# Stable rotating vortex clusters in three-dimensional quantum droplets

**Liangwei Dong[1,*], Yaroslav V. Kartashov[2]**

[1]*Department of Physics, Zhejiang University of Science and Technology, Hangzhou, 310023, China*

[2]*Institute of Spectroscopy, Russian Academy of Sciences, Troitsk, Moscow, 108840, Russia*

We predict a new type of stable three-dimensional (3D) vortex quantum droplets in binary Bose-Einstein condensate arranged into ring clusters that persistently rotate in the external potential. In contrast to clusters composed from localized quantum droplets, the states introduced here represent interacting vortex lines with identical topological charges nested in common 3D envelope and existing in much broader parameter range in comparison with localized quantum droplets. The intricate interplay between mean-field nonlinearity of Bose-Einstein condensate, quantum fluctuations described by Lee-Huang-Yang correction, Coriolis force arising due to rotation, and external potential leads to substantial variations of cluster shape upon increase of its rotation frequency. Rotating vortex clusters bifurcate from single-vortex quantum droplets and exist only above critical value of the rotation frequency that decreases with increase of chemical potential and depends on the number of vortex lines in the cluster. The radius of the cluster decreases with increase of rotation frequency and weakly varies with chemical potential, which determines mostly localization of the individual vortices in cluster and overall width of its envelope. Vortex droplet clusters are very robust objects existing in stable form in wide intervals of the number of particles for any number (odd or even) of vortex lines forming them. Our results may open the route to observation of stable arrays of vortex lines of different configurations performing regular collective motion in condensate.

## 1. Introduction

Exploration of new physical systems and mechanisms allowing the formation and observation of stable three-dimensional (3D) self-sustained localized states in nonlinear materials represents a fundamental problem attracting considerable attention in different areas of physics including optics, hydrodynamics, physics of plasmas, physics of matter waves, and optoelectronic systems [1-11]. Particularly intriguing and interesting for potential applications are multidimensional self-sustained states carrying vortices – topological defects, around which phase of the wavefunction or field winds by integer multiples of $2\pi$. Such states appear in variety of forms even in 2D geometries, including complexes of several interacting vortex solitons, while in 3D geometries self-sustained vortex solitons are characterized by the presence of vortex lines, rings, spirals or knots [12-14]. Vortices are topologically stable objects, but self-sustained states carrying them represent higher-order, excited states of the system and on this reason they are usually prone to perturbations and may be dynamically unstable [15-17]. Even when the simplest fundamental multidimensional solitons in a given system are stable (possible stability mechanisms are described in [1-11]), vortex-carrying nonlinear states in the same material may be strongly unstable, especially when nonlinearity supporting them is focusing (attractive) [12-14]. Naturally, stabilization of 3D vortex-carrying states is harder than stabilization of 2D ones, due to richer spectrum of possible perturbations in 3D systems [1-3].

Multidimensional vortex-carrying states have been studied in various optical and matter-wave systems. They can be created using trapping potentials, including harmonic-oscillator [18-21] and periodic ones [22,23]. While the former potentials allowed the observation of stable vortex states for the case of dominating repulsive nonlinearity [24-30], the latter ones can support stable 2D [31-35] and potentially 3D [36-43] vortex solitons even for attractive nonlinearity. Nonlocal [44,45] or nonuniform [46-50] nonlinearities can also give rise to stable 2D and 3D vortex solitons. Spin-orbit coupling in Bose-Einstein condensates [51,52] allows to create stable 2D [53-56] and 3D [57,58] self-sustained states, where vorticity naturally emerges in one of the condensate components as a result of spin-orbit coupling. Dissipative systems with competing gain and nonlinear losses [59,60] may also support stable 3D vortex solitons. However, despite the predictions of stable 3D self-sustained vortex states, they have not been observed in attractive BECs, while in optics they were observed only as transient, actually unstable objects [43].

Recent discovery of quantum droplets – self-bound liquid-like states in Bose-Einstein condensate stabilized by the balance between mean-field attraction and quantum fluctuations described by the Lee-Huang-Yang (LHY) correction [61,62] – has opened particularly broad prospects for the experimental realization of stable 3D self-sustained states. Quantum droplets in their simplest, fundamental form have been observed in dipolar [63-65,29,30] and binary [66-70] condensates. In the latter case, the stability of quantum droplets is secured by the repulsive quantum-mechanical correction to the effectively attractive mean-field energy, which is a residue of the nearly exact balance of the inter- and intra-species interactions in binary BEC with opposite signs. This stabilization mechanism, similar to the mechanism of competing nonlinearities, has allowed the prediction of stable quantum droplets of different types, see reviews [71-73]. Remarkably, it also predicts the existence of stable 2D [74-77] and 3D [78-81] single-vortex quantum droplets, both in free space and in the presence of the external potentials [82-85]. Complex 3D topological states, such as hopfions [86,87] can also be realized in quantum droplets.

Importantly, when condensate is set into rotation, its properties may drastically change [13]. Thus, rotation may be accompanied by the emergence of stable clusters of several co-rotating vortices, and, for fast rotations, of vortex lattices, whose properties are well understood for confined 2D BECs with repulsive mean-field nonlinearity [13,25,88-96]. Very recently it was shown that clusters of persistently rotating localized quantum droplets are possible [97,98], as well as clusters of vortices nested in them [99,100] (even in the regime of fast rotation [101-103]). However, such vortex clusters were considered exclusively in 2D configurations, and were not reported in 3D ones.

In this paper we predict the existence of stable 3D vortex droplet clusters composed of several co-rotating vortices (vortex lines) with the same topological charge located on a ring and nested in common droplet envelope. Vortex clusters are stabilized by the repulsive Lee-Huang-Yang quantum corrections, since the residual mean-field cubic nonlinearity in this system is *attractive*. Vortex clusters emerge only above critical rotation frequency and their existence is tightly

*Corresponding author's email: dlw_0@163.com

linked to the instability of 3D single-vortex quantum droplets, from which such clusters bifurcate. The distance between vortices in cluster decreases with increase of the rotation frequency. Our results may open the route to experimental realization of stable 3D vortex clusters performing controllable collective motion.

## 2. Theoretical model

We consider binary BEC in the external potential in 3D geometry, where the evolution of the wavefunctions $\psi_{1,2}$ of two components is governed by the coupled dimensionless Gross-Pitaevskii equations [9,61,66,68]:

$$i\frac{\partial\psi_{1,2}}{\partial t}=-\frac{1}{2}\Delta_{\rm 3D}\psi_{1,2}+(|\psi_{1,2}|^2-g|\psi_{2,1}|^2)\psi_{1,2}+ g_{\rm LHY}|\psi_{1,2}|^3\psi_{1,2}-\mathcal{V}(x,y,z)\psi_{1,2}, \qquad (1)$$

where $\Delta_{\rm 3D}=(\partial_x^2+\partial_y^2+\partial_z^2)$, $t$ is the evolution time, the strength of intra-species repulsion is normalized to $1$, $g>1$ is the strength of inter-species attraction, the Lee-Huang-Yang quantum correction leads to additional quartic self-repulsion $\sim g_{\rm LHY}$, and we initially consider super-Gaussian potential $\mathcal{V}=pe^{-(x^2+y^2)^4/d_r^8-z^8/d_z^8}$ with depth $p$, radial $d_r$ and axial $d_z$ widths. Here we use normalizations adopted in [78,104], assuming for simplicity the species with equal masses and intra-species interactions. Scattering lengths defining $g, g_{\rm LHY}$ values can be tuned by means of the Feshbach resonance [66,68]. We set $g=1.12$ (i.e. residual mean-field nonlinearity is attractive), $g_{\rm LHY}=0.135$, while the parameters of potential are set as $d_r=13$, $d_z=2$, $p=6$. This ensures that the ratio of axial/radial BEC widths for low densities is approximately $1:6$, i.e. such states are truly 3D (for examples of states with $1:1$ width ratio, see Fig. 5). Among the conserved quantities of Eq. (1) is the total norm

$$N=\iiint(|\psi_1|^2+|\psi_2|^2)dxdydz$$

defining number of particles in BEC and energy

$$\mathcal{E}=(1/2)\iiint[|\nabla\psi_1|^2+|\nabla\psi_2|^2+|\psi_1|^4+|\psi_2|^4- 2g|\psi_1|^2|\psi_2|^2+(4/5)g_{\rm LHY}(|\psi_1|^5+|\psi_2|^5)- \mathcal{V}(|\psi_1|^2+|\psi_2|^2)]dxdydz.$$

In normalizations adopted here the dimensionless norm $N$ is connected with number of atoms $\mathcal{N}$ in the condensate as $\mathcal{N}=(\varepsilon_0 r_0^3/g_s)N$. For characteristic transverse scale of $r_0=0.5\ \mu{\rm m}$, one gets energy and unit time scales of $\varepsilon_0\sim 6.9\times10^{-31}$ J and $t_0\sim 0.15$ ms (in these estimates we used atomic mass of $^{39}$K from recent experiments [66,67]). For simplicity, we search for droplet clusters with identical functional form of two components $\psi_1=\psi_2=\psi$ (as it may potentially occur in homonuclear mixtures [68]), representing stationary states in the coordinate frame $x'=x\cos(\omega t)+y\sin(\omega t)$, $y'=y\cos(\omega t)-x\sin(\omega t)$, $z'=z$ rotating with frequency $\omega$ in the $(x,y)$ plane, where the wavefunction $\psi$ satisfies the equation (further we omit primes in coordinates):

$$i\frac{\partial\psi}{\partial t}=-\frac{1}{2}\Delta_{\rm 3D}\psi+(1-g)|\psi|^2\psi+g_{\rm LHY}|\psi|^3\psi-\mathcal{V}\psi+ i\omega(x\partial_y\psi-y\partial_x\psi). \qquad (2)$$

The last term in the right-hand side of Eq. (2) accounts for Coriolis force affecting self-sustained states in the rotating frame (here positive $\omega$ corresponds to counterclockwise rotation). Stationary vortex droplet clusters $\psi=we^{-i\mu t}$ [here $w(x,y,z)$ is the complex function describing cluster shape] were found using Newton method.

## 3. Results and discussions

The examples of 3D vortex clusters comprising two, three, and four co-rotating vortex lines are depicted in Fig. 1 together with representative $N(\mu)$ dependencies [red lines in Fig. 1(a),(c), and (e)] for fixed $\omega=0.08$. To illustrate the structure of vortex cluster, we show 3D isosurfaces of $|\psi|$ at two different levels, as well as modulus $|\psi|$ and phase $\arg(\psi)$ distributions in $z=0$ plane. These states are clearly distinct from previously studied rotating *localized* quantum droplets, since they are composed of vortex lines (each has the same topological charge $m=1$) nested in common envelope. At fixed $z$ all vortices are located equidistantly on a ring of certain radius. Typically the distance between centers of adjacent vortices increases with increase of the number of vortices $n$ in cluster. Families of vortex droplet clusters bifurcate from the families of single-vortex quantum droplets of the form $\psi=w(r,z)e^{im\phi-i\mu t}$ [here $m$ is the topological

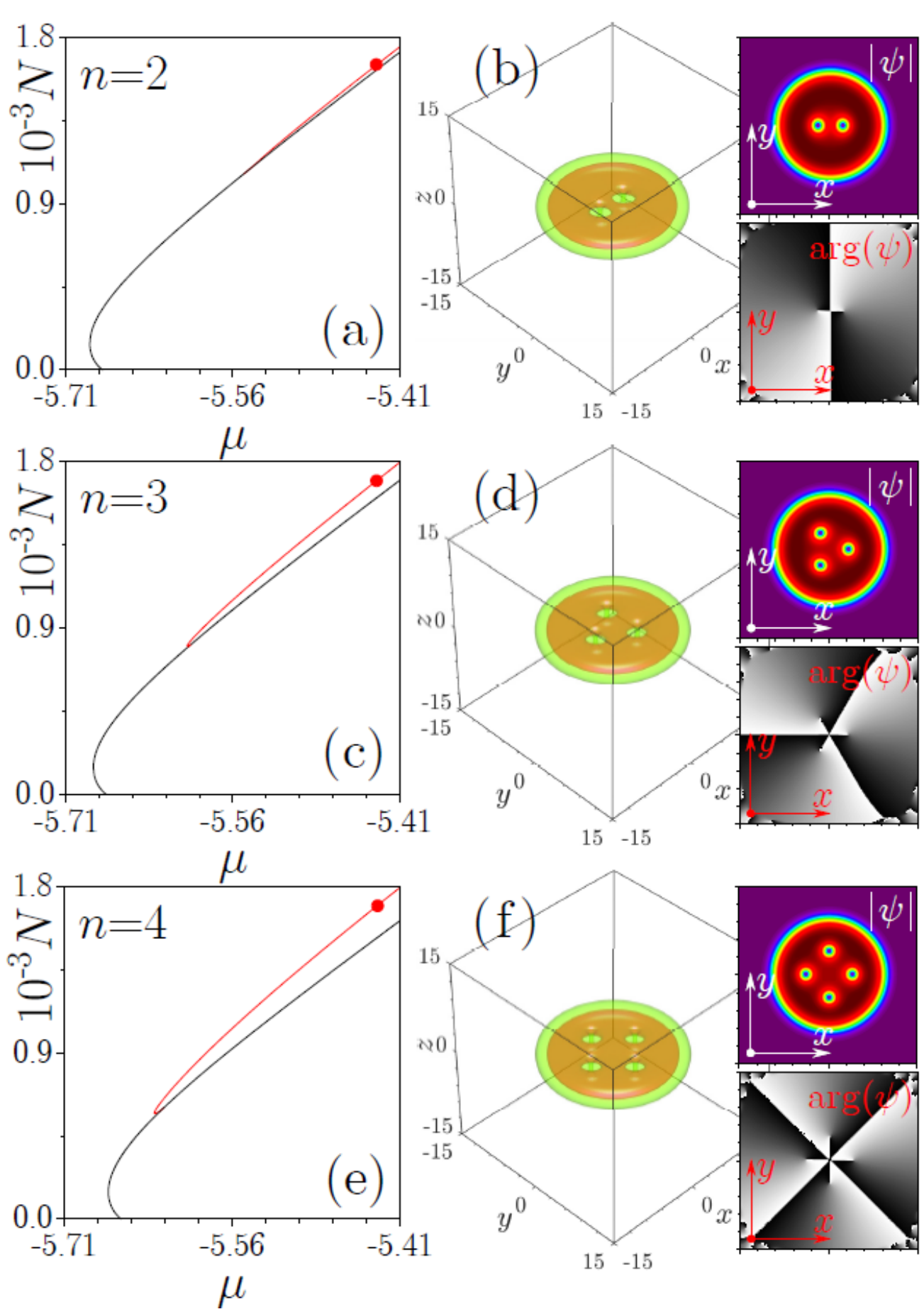

Fig. 1. Dependence of norm of 3D vortex droplet cluster containing $n=2$ (a), 3 (c), and 4 (e) vortex lines on chemical potential $\mu$ (red curves). Black curves show $N(\mu)$ dependencies for single-vortex droplet with topological charge $m=2$ (a), 3 (c), and 4 (e), from which clusters bifurcate. Panels (b),(d),and (f) show examples of vortex droplet clusters at $\mu=-5.43$ corresponding to the red dots in panels (a), (d), and (e), respectively. Two superimposed isosurfaces of $|\psi|$ are shown that correspond to $0.2\max|\psi|$ (green)

and $0.7\max|\psi|$ (orange) levels. Insets show modulus $|\psi|$ and phase $\arg(\psi)$ distributions in the $(x,y)$ plane at $z=0$. In all cases rotation frequency $\omega=0.08$. Notice that the families of vortex clusters and their stability were analyzed in much wider intervals of $\mu$ in comparison with those shown in Fig. 1(a),(c),(e).

charge of single central vortex line, $(r,\phi)$ are radius and polar angle in the $(x,y)$ plane], shown with black lines in Fig. 1(a),(c), and (e). The number $n$ of vortices in the cluster coincides with topological charge $m$ of single-vortex state from which cluster emanates. Thus, vortex clusters exist only above critical norm $N_{\mathrm{bf}}$ (the point where red curves split from black ones) that decreases with $n$ for fixed $\omega$. The bifurcation point $\mu=\mu_{\mathrm{bf}}$ shifts to the right as $\omega$ decreases leading to fast increase (and eventual divergence) of $N_{\mathrm{bf}}$, i.e. the rotation is necessary for the existence of vortex droplet clusters.

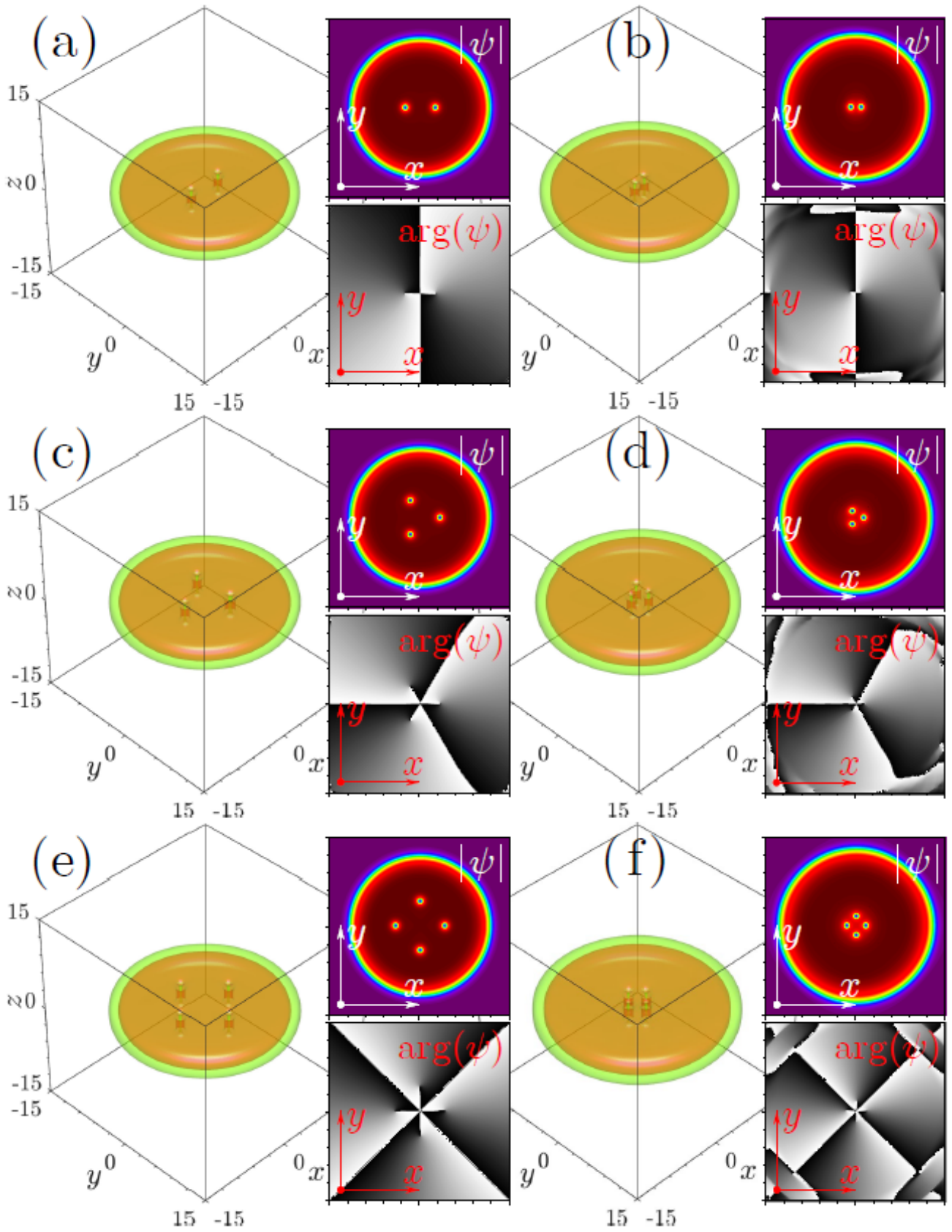

Fig. 2. Examples of rotating clusters of two (a),(b), three (c),(d), and four (e),(f) vortex droplets at $\mu=-3.5$. Panels (a),(c),(e) correspond to rotation frequency $\omega=0.08$, while panels (b),(d),(f) correspond to $\omega=0.50$. Insets show modulus $|\psi|$ and phase $\arg(\psi)$ distributions in the $(x,y)$ plane at $z=0$. The superimposed isosurfaces of $|\psi|$ for each state are drawn at $0.2\max|\psi|$ (green) and $0.7\max|\psi|$ (orange) levels.

Single-vortex droplets, from which clusters bifurcate, in turn emerge from linear eigenstates of the potential $\mathcal{V}$ in rotating coordinate frame that can be obtained from linearized version of Eq. (2). The eigenvalues of these linear single-vortex states ($\mu_{m=2}\approx-5.677$, $\mu_{m=3}\approx-5.674$, and $\mu_{m=4}\approx-5.659$) correspond to the points in Fig. 1 where norm of single-vortex state vanishes. Since Coriolis term $\omega(x\partial_y\psi-y\partial_x\psi)$ in Eq. (2) can be written as $\omega\partial_\phi\psi$ in $(r,\phi,z)$ coordinates, the substitution $\psi=w(r,z)e^{im\phi-i\mu t}$ (valid only for radially-symmetric states) yields the equation for $w(r,z)$ with $\mu+m\omega$ replacing chemical potential $\mu$ in nonrotating frame. This means that for $m,\omega>0$ the rotation simply shifts the dependencies $N(\mu)$ for radially-symmetric states by $m\omega$ in the negative direction of the $\mu$ axis. As one can see from Fig. 1, the norm of single-vortex state initially increases with decrease of $\mu$ (in this interval residual cubic self-attraction dominates over repulsive LHY term), up to the turning point $\mu_m^{\mathrm{turn}}$, where critical amplitude is reached that can be roughly estimated as $w_{\mathrm{turn}}\sim(g-1)/g_{\mathrm{LHY}}$ (for amplitudes above this value the LHY term becomes dominating). Further increase of amplitude leads to turn of $N(\mu)$ curve and progressively increasing expansion of single-vortex droplets as $\mu\to0$. Vortex clusters bifurcate from single-vortex droplets namely in this regime, where LHY correction becomes dominating.

As $\mu$ increases, the overall width of the cluster envelope increases (in both radial and axial directions) and it eventually acquires flat-top shape (see Fig. 2, where envelopes for larger values of $\mu$ are clearly wider than in Fig. 1) that results in divergence of norm $N$ when $\mu\to0$. Instead, the widths of individual vortices nested in envelope decrease with $\mu$. The existence domain $\mu_{\mathrm{bf}}(\omega)\le\mu<0$ of the vortex droplet clusters found here is substantially wider than the domain of existence $-(25/216)(g-1)^3g_{\mathrm{LHY}}^{-2}\le\mu<0$ of localized quantum droplets in free space [78].

Rotation significantly affects shapes of vortex clusters. As Fig. 2 shows, the increase of the rotation frequency $\omega$ at fixed chemical potential $\mu$ leads to the reduction of the cluster radius (and the distance $S$ between adjacent vortex lines), while the overall width of the envelope, where vortex lines are nested, instead increases. This behavior is observed for any number (odd or even) of vortex lines $n$ in cluster (Fig. 2). The norm of the cluster monotonically increases with $\omega$ [Fig. 3(a)]. Notice that the steepness of $N(\omega)$ dependence increases with increase of the number of vortices in cluster. It should be mentioned that for negative rotation frequencies $\omega$ one can obtain similar dependencies, with the only difference that clusters will be composed from $n$ vortices with charge $-1$ (bifurcating from single-vortex states with $m=-n<0$).

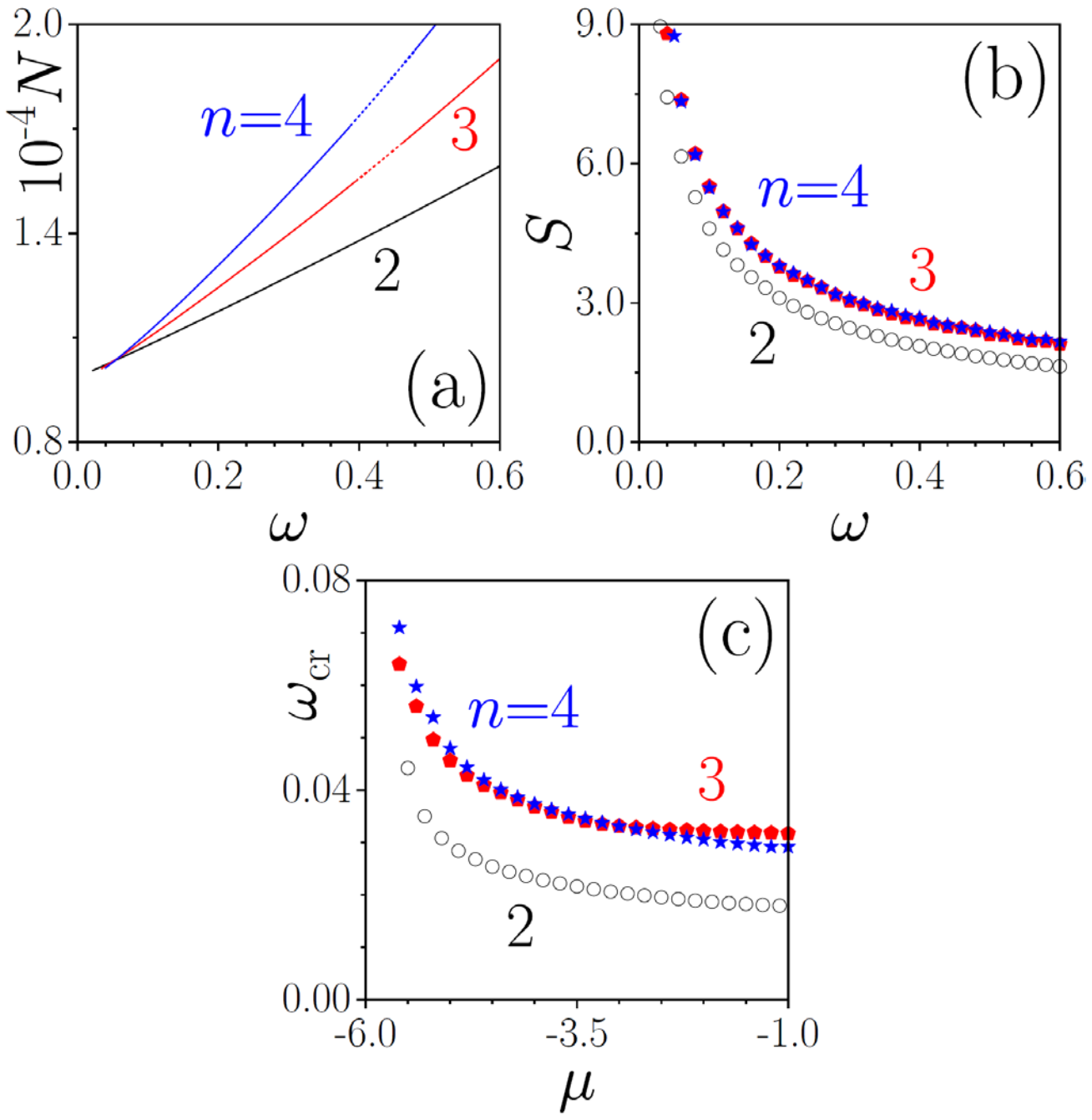

Fig. 3. (a) Norm $N$ of the vortex droplet clusters with different number of vortices $n$ and (b) the distance $S$ between adjacent phase singularities versus rotation frequency $\omega$ at $\mu=-3.5$. In (a) solid

lines correspond to stable clusters, while dotted to unstable ones. (c) Critical rotation frequency $\omega_{\rm cr}$ of vortex droplet clusters vs chemical potential $\mu$.

The dependence of the distance $S$ between adjacent singularities (calculated in $z=0$ plane) on rotation frequency $\omega$ at fixed $\mu$ is presented in Fig. 3(b). It shows that vortex cluster broadens with decrease of $\omega$ for all values of $n$. This figure reveals a remarkable fact that at a given $\mu$ vortex droplet clusters can exist only when the rotation frequency exceeds certain critical value $\omega_{\rm cr}$ that depends on $n$. For example, at $\mu=-3.5$ the critical rotation frequencies are $\omega_{\rm cr}^{n=2}\approx 0.0216$, $\omega_{\rm cr}^{n=3}\approx 0.0344$, and $\omega_{\rm cr}^{n=3}\approx 0.0398$. Since when $\omega\to\omega_{\rm cr}$ the radius of the cluster becomes comparable with the width of potential in the $(x,y)$ plane (i.e. vortex lines move to its periphery), we believe the existence of critical frequency is due to finite width of potential $\mathcal{V}$. The connection between critical rotation frequency and finite width of potential can be elucidated using the approach suggested in [105,106] for 2D vortex polygons. The critical rotation frequency monotonically decreases with increase of $\mu$ [Fig. 3(c)]. Notice that $\omega_{\rm cr}$ is lower for cluster with $n=2$ vortices, while for $n=3,4$ the critical frequencies are close. In addition to simple clusters with $n=2-4$ discussed above, we found rotating vortex clusters with $n$ up to $8$, as well as clusters containing simultaneously one central and several off-center vortices with the same topological charge $+1$. The examples of such structures are shown in Fig. 4(a) and 4(b). The critical rotation frequency, above which such clusters form, is larger than $\omega_{\rm cr}$ for clusters with fewer vortices.

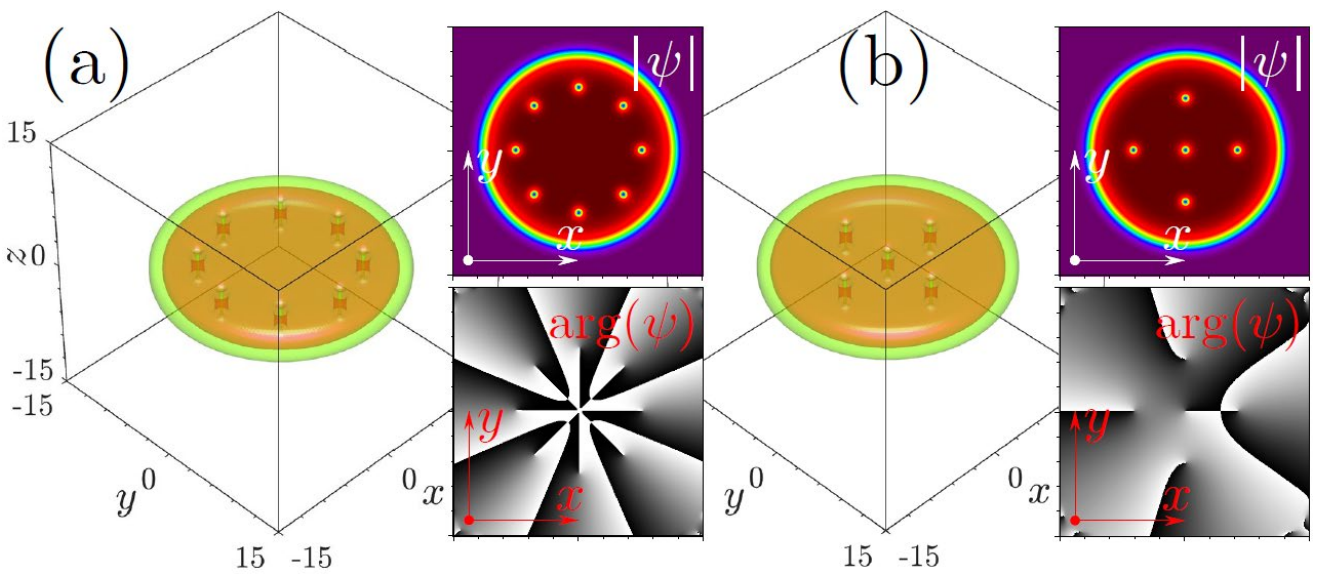

Fig. 4. Examples of rotating vortex clusters with 8 vortices (a) and with 4 off-center and 1 central vortex (b) at $\mu=-3.5$, $\omega=0.06$.

Importantly, rotating vortex clusters can be obtained in a wide variety of 3D potentials, including spherically symmetric ones, i.e. $\mathcal{V}=pe^{-(x^2+y^2+z^2)^4/d_\rho^8}$, where $p=6$, $d_\rho=13$. The examples of 3D clusters of co-rotating vortices in such potential are shown in Fig. 5. In this geometry vortex lines may be notably curved. Their curvature increases with decrease of rotation frequency and they clearly move toward the periphery of potential as $\omega\to\omega_{\rm cr}$. Instead, for large frequencies vortex lines become more straight and shift toward the center of the common spherical envelope. In contrast to vortex droplet clusters in super-Gaussian potential, the instability domains for vortex droplet clusters in spherically symmetric potential are relatively narrow and are located for a given value of $\mu$ at rotation frequencies $\omega$ close to $\omega_{\rm cr}$ value. For example, vortex clusters with two vortex lines are unstable in the region $\omega_{\rm cr}=0.04<\omega<0.056$. The instability region shrinks with the increase of the number of vortex lines $n$ in cluster.

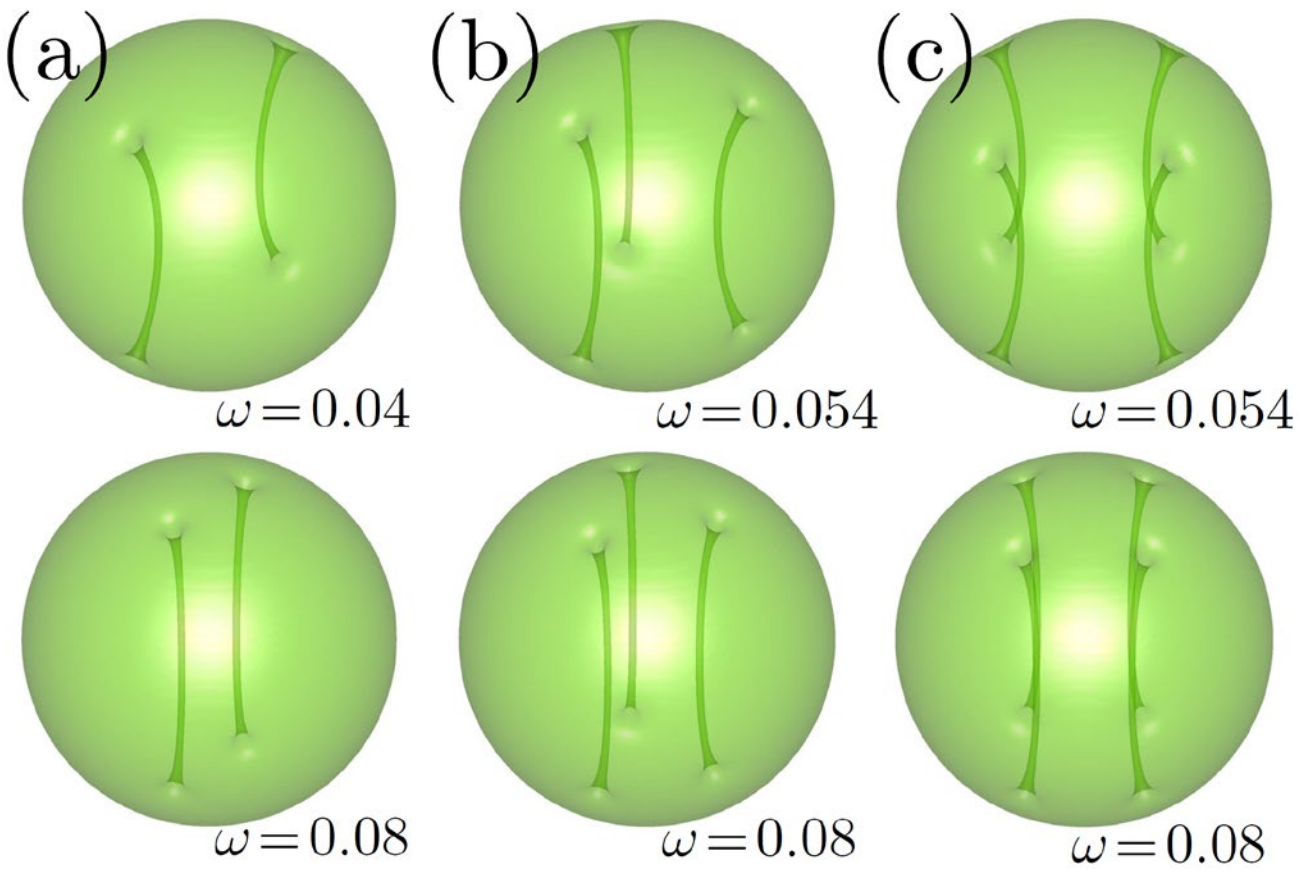

Fig. 5. Isosurfaces drawn at $0.35\max|\psi|$ level showing rotating clusters of two (a), three (b), and four (c) quantum droplets in spherical potential at $\mu=-3.5$ and different rotation frequencies.

Rotating vortex clusters are very robust objects. To test their stability, we added small perturbations into initial wavefunction $\psi|_{t=0}$ describing cluster and modeled its evolution over large times $t_{\rm evol}=10^4$ that allows to capture even weak instabilities (even with growth rates $\lambda_{\rm re}\sim 1/t_{\rm evol}$), if they can develop. Note that such stability analysis based on long-time direct numerical propagation is quite time-consuming and sometimes stability borders defined with this approach may be defined with finite accuracy, but it usually allows to clearly distinguish different types of instabilities. One nevertheless has to resort to this approach in 3D problems for self-sustained states that do not feature radial or spherical symmetries, since rigorous Bogoliubov-de Gennes stability analysis would be unfeasible due to huge size of the matrices arising in resulting linear eigenvalue problem. In super-Gaussian potential $\mathcal{V}=pe^{-(x^2+y^2)^4/d_r^8-z^8/d_z^8}$ the rotating clusters of two vortices at $\mu=-3.5$ are stable in their entire existence domain in $\omega$. The example of stable rotation with period $T=2\pi/\omega$ of such cluster (in the non-rotating coordinate frame) is shown in Fig. 6(a). For the same $\mu$ the three and four-vortex clusters are unstable in narrow frequency intervals $\omega\in[0.396,0.458]$ and $[0.388,0.478]$, respectively. Corresponding stable branches are shown with solid lines in Fig. 3(a), while unstable ones are shown with dotted lines. The instability arises as a pitchfork bifurcation at sufficiently large values of $\omega$, similar to the instability occurring for vortex clusters trapped in a 2D harmonic potential [96]. Families of vortex clusters depicted in Fig. 1 are stable at least within shown domains near bifurcation points in $\mu$. The example of stable rotation of three-vortex cluster is shown in Fig. 6(b), while instability development for three- and four-vortex clusters is shown in Fig. 6(c) and 6(d). It should be mentioned that upon instability development the individual vortex lines experience irregular displacements, but they are not destroyed upon long-time evolution and remain near the center of the common envelope. We found that for more complex vortex clusters with $n>4$, the stability region quickly shrinks with the growth of $n$. While clusters with one central and four off-center vortex lines evolve stably in relatively broad region of $(\mu,\omega)$ parameters [see example in Fig. 6(e)], vortex droplet clusters with $n=8$ are completely unstable.

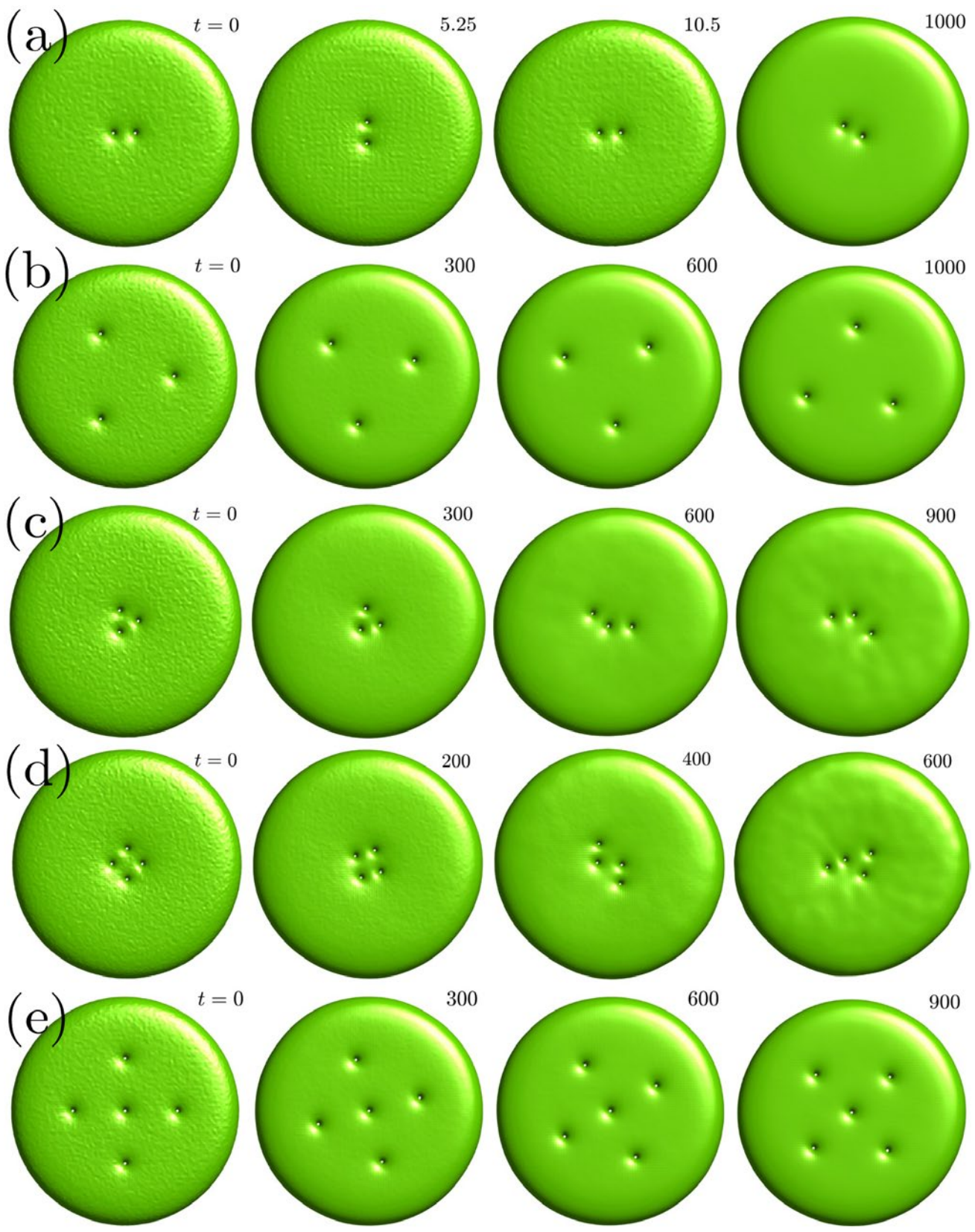

Fig. 6. Isosurfaces at $0.2 \max|\psi|$ level at different moments of time illustrating stable evolution of perturbed (a) two-vortex cluster at $\omega = 0.3$, (b) three-vortex cluster at $\omega = 0.036$, and (e) five-vortex cluster at $\omega = 0.06$. Decay of the (c) unstable three-vortex cluster at $\omega = 0.45$, and (d) unstable four-vortex cluster at $\omega = 0.43$. In all cases $\mu = -3.5$.

## 4. Conclusions

Summarizing, we introduced a new type of stable rotating 3D quantum droplet clusters, stabilized by joint action of LHY quantum corrections, potential and rotation. Our findings demonstrate that the presence of the external potential is crucial for the existence of such states that emerge only above certain critical rotation frequency and can exist in potentials with various shapes. We have shown that the existence of clusters is tightly connected with stability properties of conventional vortex quantum droplets in the rotating coordinate frame. These results may open the route to observation of stable 3D vortex-carrying states and their complexes in quantum droplets that perform regular collective motion. Similar self-sustained objects can be potentially encountered in other physical systems, including nonlinear optical materials, magnetic fluids, superconductors, not only in conservative, but also in dissipative settings.

### CRediT authorship contribution statement

L. D.: Conceptualization, Investigation; Y.V.K.: Writing – review & editing, Validation.

### Declaration of competing interest

The authors declare that they have no known competing financial interests or personal relationships that could have appeared to influence the work reported in this paper.

### Acknowledgements

L.D. acknowledges support from the National Natural Science Foundation of China (NSFC) (grant No. 62575264). Y.V.K. acknowledges support from the research project FFUU-2024-0003 of the Institute of Spectroscopy of the Russian Academy of Sciences.

### Data Availability

Data will be made available on request.